\documentclass[lettersize,journal]{IEEEtran}
\usepackage{amssymb}
\usepackage{graphicx}
\usepackage{siunitx}
\usepackage{verbatim}
\usepackage{amsmath,amsfonts}
\usepackage{algorithmic}
\usepackage{algorithm}
\usepackage{array}
\usepackage[caption=false,font=normalsize,labelfont=sf,textfont=sf]{subfig}
\usepackage{textcomp}
\usepackage{stfloats}
\usepackage{url}
\usepackage{verbatim}
\usepackage{cite}
\usepackage{hyperref}

\newcommand{\orcid}[1]{\href{#1}{\includegraphics[height=10pt]{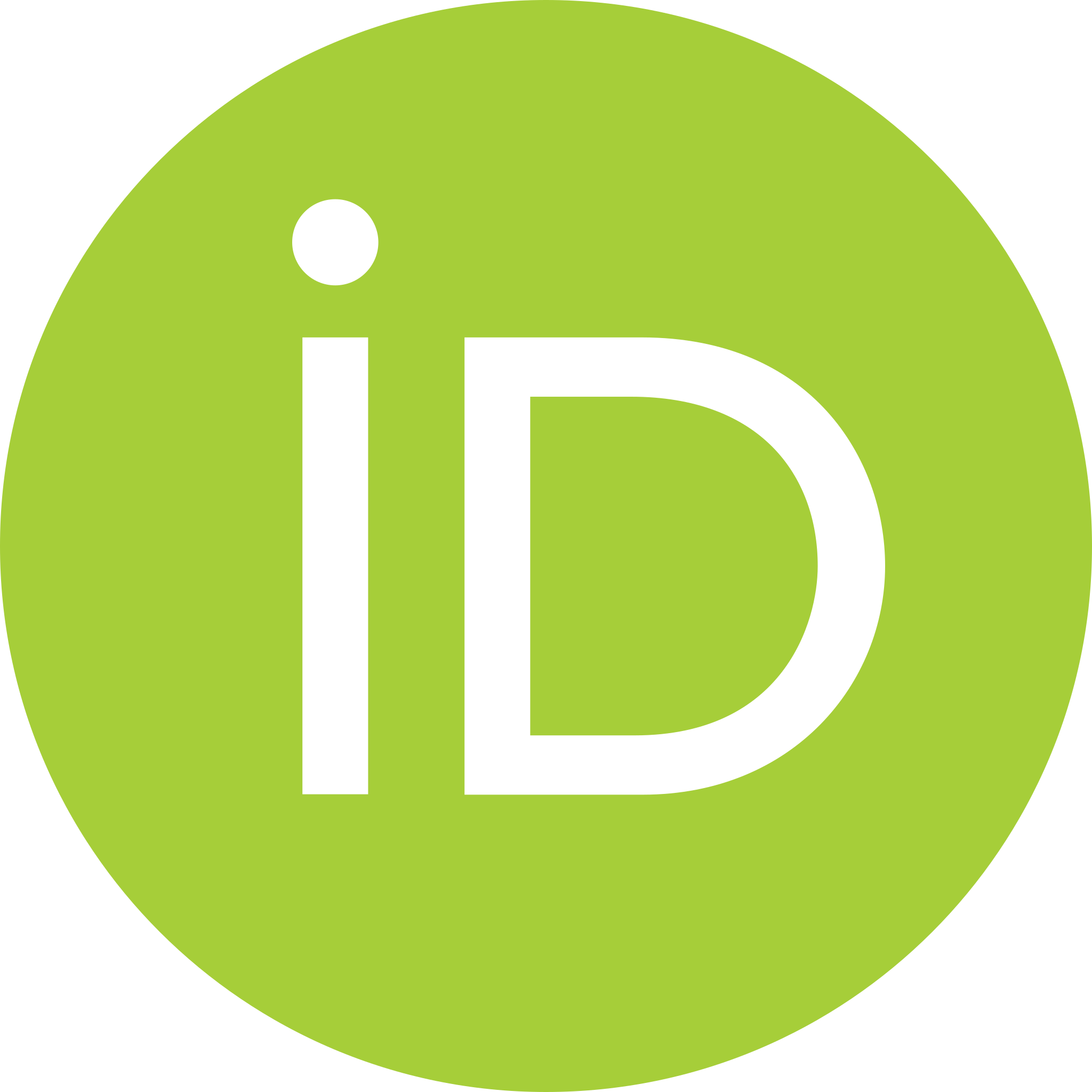}}}

\begin{document}

\title{JCO: Optimization Framework for Nonlinear Superconducting Circuits Using a Lumped-Element Approach and Harmonic Balance}

\author{E.~Palumbo \orcid{https://orcid.org/0009-0006-7155-8249},
A.~Alocco\orcid{https://orcid.org/0009-0000-3416-2620},
A.~Celotto\orcid{https://orcid.org/0009-0002-0759-6049},
L.~Fasolo \orcid{https://orcid.org/0000-0002-7537-7772},
B.~Galvano\orcid{https://orcid.org/0009-0001-3782-5618},  
P.~Livreri \orcid{https://orcid.org/0000-0001-8599-0418}, 
and~E.~Enrico \orcid{https://orcid.org/0000-0002-2125-5200}

\thanks{E.~Palumbo, A.~Alocco and A.~Celotto are with Dipartimento Scienza Applicata e Tecnologia, Politecnico di Torino, Corso Duca degli Abruzzi, 24, I-10129 Turin, Italy and INRiM - Istituto Nazionale di Ricerca Metrologica, Strada delle Cacce, I-10135 Turin, Italy.}
\thanks{L.~Fasolo and E.~Enrico are with INRiM - Istituto Nazionale di Ricerca Metrologica, Strada delle Cacce, I-10135 Turin, Italy.}
\thanks{B.~Galvano is with University of Palermo, Department of Engineering, I-90128, Palermo, Italy and INRiM - Istituto Nazionale di Ricerca Metrologica, Strada delle Cacce, I-10135 Turin, Italy.}
\thanks{P.~Livreri is with University of Palermo, Department of Engineering, I-90128, Palermo, Italy, and CNIT, RaSS, Pisa, Galleria Gerace 14, I-56124, Pisa, Italy.}
\thanks{Corresponding author: E. Enrico, e.enrico@inrim.it}
}

\maketitle

\begin{abstract}
In this contribution, we present JosephsonCircuitsOptimizer.jl (JCO), a simulation and optimization framework based on the JosephsonCircuits.jl library for Julia.
It models superconducting circuits that include Josephson junctions (JJs) and other nonlinear elements within a lumped-element approach, leveraging harmonic balance—a frequency-domain technique that provides a computationally efficient alternative to traditional time-domain simulations. 
JCO automates the evaluation of optimal circuit parameters by implementing Bayesian optimization with Gaussian processes through a device-specific metric and identifying the optimal working point to achieve a defined performance function. This makes it well-suited for circuits with strong nonlinearity and a high-dimensional set of coupled design parameters.
To demonstrate its capabilities, we focus on optimizing a Josephson Traveling-Wave Parametric Amplifier (JTWPA) based on Superconducting Nonlinear Asymmetric Inductive eLements (SNAILs), operating in the three-wave mixing (3WM) regime. The device consists of an array of unit cells, each containing a loop with multiple JJs, that amplifies weak quantum signals near the quantum noise limit.
By integrating efficient simulation and optimization strategies, the framework supports the systematic development of superconducting circuits for a broad range of applications.
\end{abstract}


\IEEEpeerreviewmaketitle

\section{Introduction}
\IEEEPARstart{S}{uperconducting} quantum circuits based on Josephson junctions (JJs) play a central role in the advancement of quantum technologies ~\cite{krantz2019, devoret2013,wendin2017}.
Their low dissipation, strong intrinsic nonlinearity, and suitability for coherent microwave control make them ideal candidates for implementing key components in quantum computing and quantum signal processing~\cite{paik2011,frattini2018, blais2004}. 

The simulation of such circuits presents significant computational challenges. Their strong nonlinearity and large number of degrees of freedom typically result in high-dimensional, nonlinear differential equations that are computationally expensive to solve. 
Traditional time-domain methods often require long integration times and fine temporal resolution, making them unsuitable for efficiently exploring circuit behavior across broad parameter spaces~\cite{peng2022, levochkina2024, maas1997}. 
In contrast, frequency-domain techniques such as harmonic balance offer a more efficient strategy by directly computing the steady-state response of the system in terms of its frequency components, without simulating its time evolution.
This approach is particularly well-suited for circuits operating under continuous, periodic driving, where capturing the long-term behavior efficiently and reliably is essential \cite{peng2022, levochkina2024}.

To enable fast and scalable simulations of nonlinear superconducting circuits, the \texttt{JosephsonCircuits.jl} library \cite{josephsoncircuits} for the Julia programming language provides a powerful and flexible framework for modeling lumped-element superconducting circuits that include JJs and other nonlinear components.
It implements harmonic balance for frequency-domain steady-state analysis \cite{maas1997}, enabling efficient exploration of a large number of circuit configurations.

This paper presents \texttt{JosephsonCircuitOptimizer.jl} (JCO)~\cite{josephsoncircuitsoptimizer}, a simulation and optimization framework built on top of \texttt{JosephsonCircuits.jl}.
Existing approaches to superconducting circuit optimization target objectives such as parameter matching through electromagnetic simulations~\cite{eriksson2025physicsguided}, robustness to fabrication variations~\cite{matsuoka2025marginx}, and optimization of quantum properties from Hamiltonian models~\cite{aumann2022circuitq, rajabzadeh, yan_qubit}, using strategies such as physics-guided updates, gradient-based methods, and data-driven surrogate models~\cite{bansal_nn}, often within broader design frameworks~\cite{fourie2023coldflux}. However, they do not explicitly address strongly nonlinear systems with many interacting elements, such as large-scale circuit arrays.

JCO instead enables automated exploration of high-dimensional parameter spaces by evaluating specific cost functions.
The framework iteratively explores the circuit configurations using efficient linear simulations to minimize a device-specific metric and identify optimal circuit parameters.
Once the optimal configuration is found, the working point of the device is determined through nonlinear simulations by maximizing a target performance, such as gain, bandwidth, or noise figure \cite{peng2022, fasolo2022}.

To demonstrate its applicability and utility, we focus on the design of a Josephson Traveling-Wave Parametric Amplifier (JTWPA) \cite{Fasolo19,Esposito2023, Malnou2020}, a nonlinear superconducting device composed of an array of JJs, designed to amplify weak quantum signals with near-quantum-limited noise performance.
In particular, we consider a Superconducting Nonlinear Asymmetric Inductive eLement (SNAIL)-based JTWPA~\cite{frattini2017} operating in the three-wave mixing (3WM) regime\cite{nilsson2023, zorin2016}.
This class of devices is particularly relevant as a test case due to the large and highly coupled parameters space, the strong intrinsic nonlinearity, and the large number of unit cells, which leads to high computational cost. 

Section~\ref{ch:2} discusses the code architecture of the framework, detailing its design, key components, and functionality. Section~\ref{ch:3} presents the use case of a SNAIL-based JTWPA, demonstrating the application of the framework to a specific superconducting quantum circuit. Finally, Section~\ref{ch:4} concludes with a summary of the framework’s capabilities and potential future developments.

\section{Code Architecture} 
\label{ch:2}

\begin{figure}[t]
\begin{center}
\includegraphics[width=0.45\textwidth]{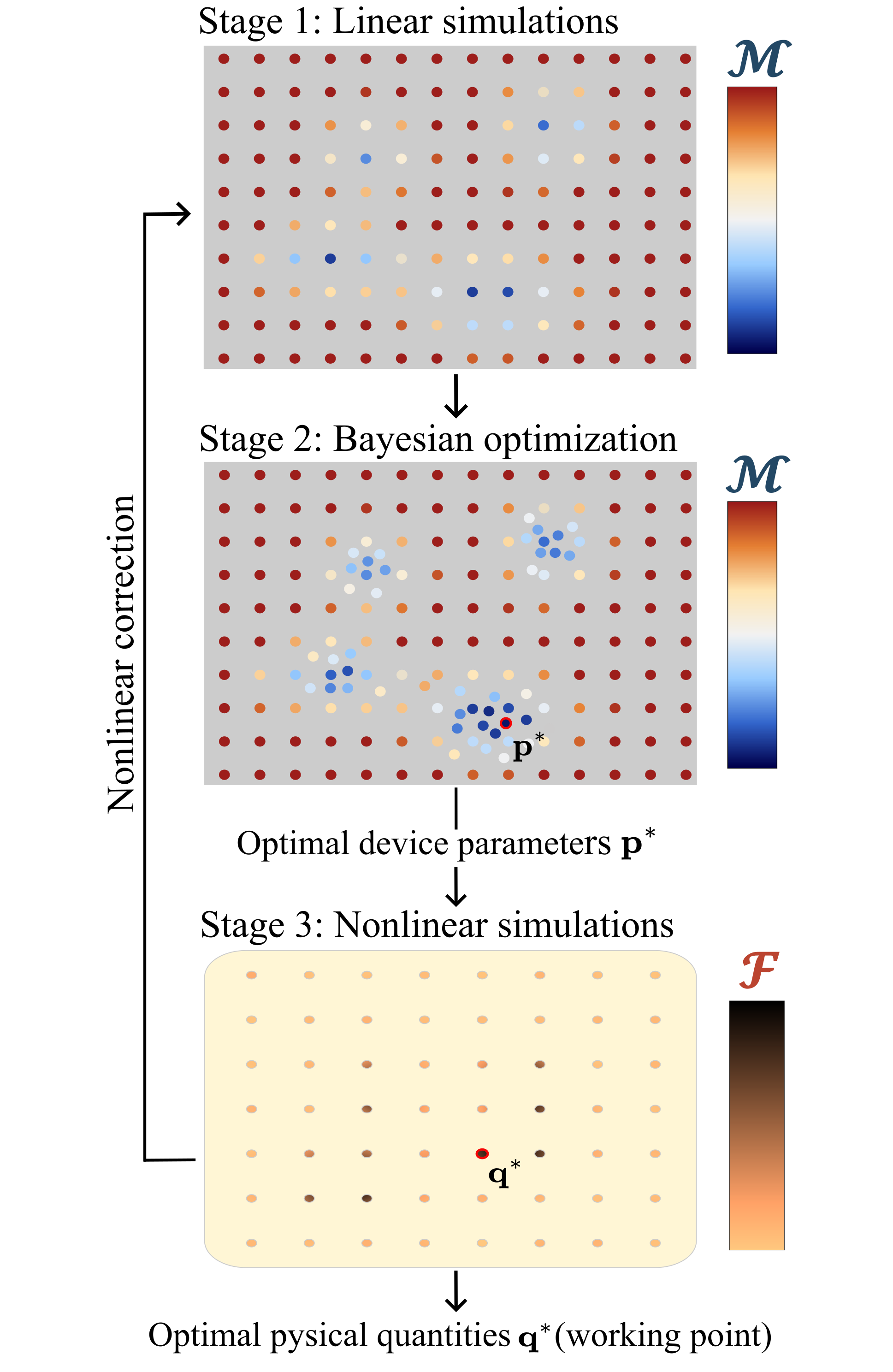}
\end{center}
\caption{Workflow of the JCO framework. The framework works as follows: (1) Linear simulations are performed across the device parameter space $\mathcal{P}$, uniformly sampled. Each configuration is assigned a metric value $\mathcal{M}$. (2) Bayesian optimization refines the sampling in regions surrounding local minima of $\mathcal{M}$ identified in the previous stage, in order to find the optimal parameters $\mathbf{p}^*$ minimizing $\mathcal{M}$. (3) Nonlinear simulations are performed by sweeping over the space of physical quantities $\mathcal{Q}$ (input frequencies and tone amplitudes) with $\mathbf{p}^*$ fixed. The process can be repeated to apply nonlinear corrections. The final outcome is the optimal device parameters $\mathbf{p}^*$ and working point $\mathbf{q}^*$ that maximizes the performance function $\mathcal{F}$. In this example, the first two stages explore two parameters taking 14 and 10 possible values, yielding $14 \times 10$ uniformly distributed configurations within $\mathcal{P}$. Each point $\mathbf{p}$ on the graph corresponds to a distinct circuit configuration, color-coded according to its metric value $\mathcal{M}$. In the third stage, two physical quantities are varied with $7 \times 8$ configurations in the $\mathcal{Q}$ space. Each point represents a different working point $\mathbf{q}$, color-coded according to the performance function $\mathcal{F}$.}
\label{fig:framework_flow}
\end{figure}

This section describes the internal structure of the JCO framework, which provides a structured pipeline that integrates circuit modeling, parametric sweeps, and Bayesian optimization into a unified and extensible workflow (Fig.~\ref{fig:framework_flow}). The JCO framework operates through a modular three-step pipeline that can be iterated to account for dynamic effects such as Kerr nonlinearity (Fig.~\ref{fig:framework_flow}).

At the foundation of the framework lies the device parameter space \( \mathcal{P} \subset \mathbb{R}^d \), defined by fabrication constraints, where \( d \) is the number of design parameters. Every point \( \mathbf{p} \in \mathcal{P} \) represents a unique circuit configuration. In our case, $d=7$ (see Section~\ref{ch:3}), and each parameter $i$ can assume $n_i$ discrete values.

\begin{figure}[t]
\begin{center}
\includegraphics[width=0.45\textwidth]{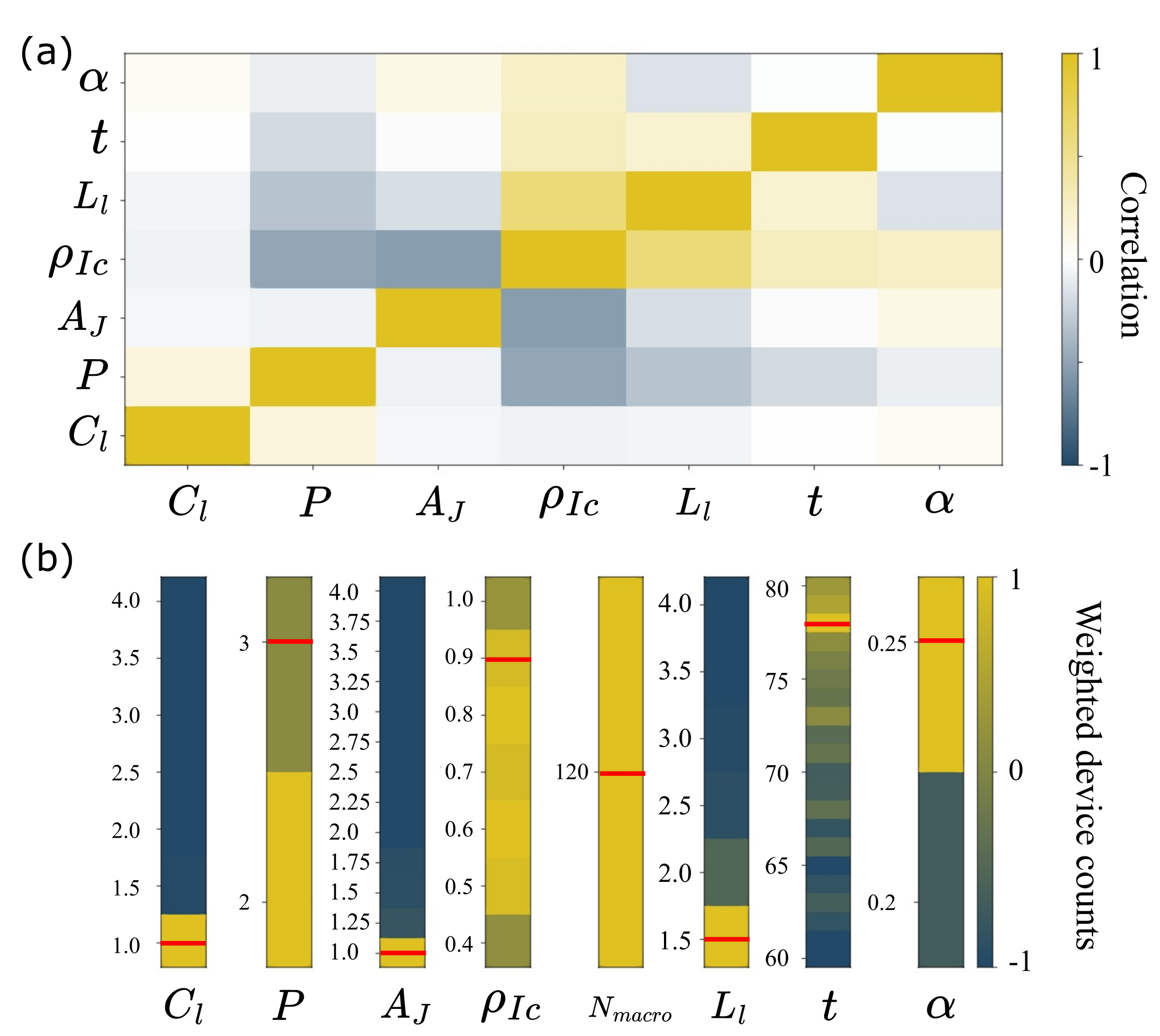}
\end{center}
\caption{(a) Correlation matrix of the device parameters and (b) 1D histogram of the parameter count weighted by the device-specific metric, for the use case presented in Section~\ref{ch:3} in a larger device parameter space. The description of each parameter is provided in Fig.~\ref{fig:snail-scheme}. The correlation matrix is computed over a subset of configurations filtered by a metric below a specified cutoff. These visualizations highlight parameter inter-dependencies and indicate which values contribute most to optimal circuit performance. After the optimization stage, the red lines define the best parameter configuration.}
\label{fig:cormat}
\end{figure}

\textbf{Stage 1 – Linear simulations.} Linear simulations are performed for a grid of uniformly sampled parameter points in \( \mathcal{P} \). Each configuration \( \mathbf{p} \) is characterized by its S-parameters $S(\mathbf{p})$, fully describing the frequency-domain behavior in the linear regime. These are computed over a broadband range—from near-DC to tens of GHz—with a resolution of about 10~MHz, depending on the desired accuracy. Each simulation is relatively fast ($\sim 2$~s) and scales with circuit complexity and frequency resolution. A device-specific metric \( \mathcal{M}(S(\mathbf{p})) \) is defined to quantify the circuit's behavior and guide the subsequent optimization. Optionally, a metric cutoff can filter configurations, highlighting relevant regions for visualization and guiding the optimization. The resulting data can be visualized through correlation matrices and one-dimensional parameter plots, revealing the design landscape and inter-dependencies (Fig.~\ref{fig:cormat}).

\textbf{Stage 2 – Bayesian optimization.} A Bayesian optimization strategy~\cite{garnett2023}, with a Gaussian Process (GPs)~\cite{Forrester2008} surrogate model, is employed to minimize \( \mathcal{M}(S(\mathbf{p})) \) over \( \mathcal{P} \), overcoming the limitations of uniform grid sampling by adaptively exploring the continuous design space. The algorithm focuses on regions surrounding local minima of $\mathcal{M}$, efficiently refining the search. This stage converges to the optimal configuration
\[
\mathbf{p}^* = \arg\min_{\mathbf{p} \in \mathcal{P}} \mathcal{M}(S(\mathbf{p})).
\]

\textbf{Stage 3 – Nonlinear simulations.} With $\mathbf{p}^*$ fixed, nonlinear simulations are performed by sweeping over a space \( \mathcal{Q} \subset \mathbb{R}^k \) of $k$ physical drive quantities, such as input frequencies and tone amplitudes. The system’s response is characterized by its X-parameters $X(\mathbf{q}; \mathbf{p}^*)$, which generalize S-parameters under strong tone conditions~\cite{root2013, peng2022}. A performance function \( \mathcal{F}(X(\mathbf{q}; \mathbf{p}^*)) \) is defined to evaluate relevant figures of merit—such as gain, bandwidth, or quantum efficiency—depending on the application~\cite{peng2022, peng2024}. These simulations are computationally heavier, typically requiring minutes, depending on circuit complexity and frequency resolution. After this stage, the framework can apply a nonlinear correction by updating the metric \( \mathcal{M} \) to include nonlinear effects (Kerr effect) and repeating the three-step cycle as needed. The final result consists of the optimal parameter set \( \mathbf{p}^* \in \mathcal{P} \) and optimal working point \( \mathbf{q}^* \in \mathcal{Q} \),
\[
\mathbf{q}^* = \arg\max_{\mathbf{q} \in \mathcal{Q}} \mathcal{F}(X(\mathbf{q}; \mathbf{p}^*)),
\]
which defines the best operating conditions for the selected circuit design.

In summary, the framework separates the process into two hierarchical stages: design-space optimization over \( \mathcal{P} \) using \( \mathcal{M} \), and working point selection over \( \mathcal{Q} \) using \( \mathcal{F} \), ensuring systematic and efficient device optimization.

\section{Use Case: a SNAIL-based JTWPA} \label{ch:3}

To demonstrate the usefulness and practical relevance of the JCO framework, we consider a SNAIL-based JTWPA operating in the 3WM regime. This use case highlights how the framework adapts to a known and physically relevant problem, offering an efficient exploration of complex design spaces and identifying optimal device configurations. 

The device consists of a transmission line composed of \( N \) periodically repeated SNAIL cells \cite{frattini2017}, with in our case \( N =360\) \cite{zorin2016}. The computational cost of the simulation scales with $N$, as it sets the size of the underlying system. The periodic structure allows engineering of bandgaps and phase-matching conditions required for 3WM amplification.

\begin{figure}[!t]
\begin{center}
\includegraphics[width=2.5in]{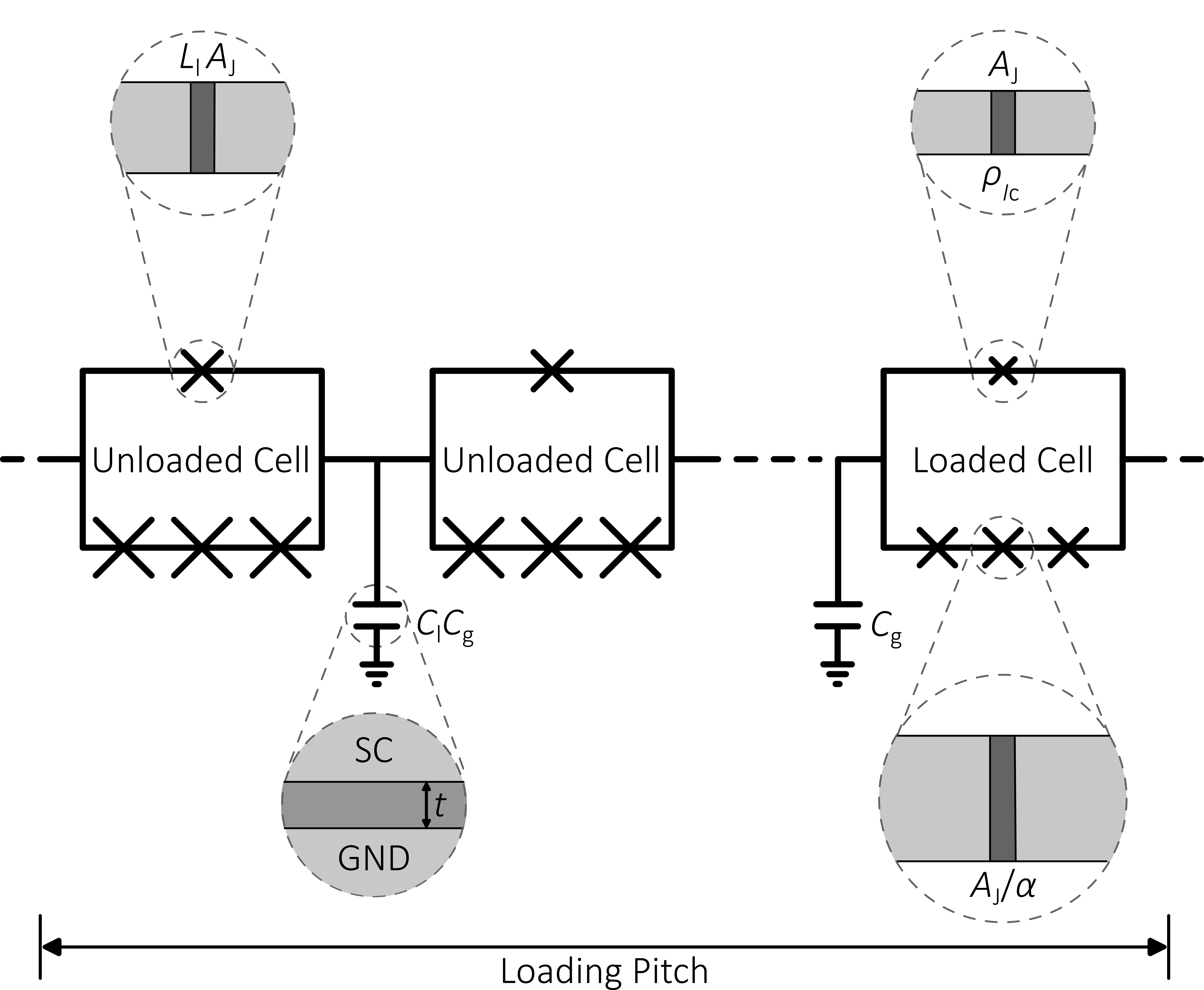}
\end{center}
\caption{Schematic of a SNAIL-based JTWPA. The device is composed of $N = N_{\text{macro}} \times P$ cells arranged in a 1D chain. Each macrocell $N_{\text{macro}}$ consists of multiple unit cells: \( P - 1 \) unloaded cells and one loaded cell, where \( P \) denotes the loading pitch. Each unit cell includes a SNAIL loop formed by two parallel branches. The first branch hosts a single small JJ with area \( A_J \) and critical current density \( \rho_{Ic} \), yielding a critical current \( I_c = \rho_{Ic} A_J \). The second branch contains three larger JJs, each with area \( A_J/\alpha \), where \( \alpha \in (0,1) \) is the asymmetry parameter. The loop is threaded by an external DC magnetic flux \( \Phi_{\mathrm{ext}} \), delivered via a dedicated flux line, which tunes the effective nonlinearity of the structure. Each cell is shunted to ground via a gate capacitance \( C_g \), determined by the dielectric thickness \( t \). Loaded and unloaded cells differ in their inductance and capacitance to modify the impedance and dispersion properties of the device. Particularly, \( L_\ell \) is the ratio between the inductance of the small junction inductance of the loaded and unloaded cell, which is proportional to the ratio between the small junction areas. \( C_\ell \) is the ratio between the ground capacitances of the loaded and unloaded cell.}
\label{fig:snail-scheme}
\end{figure}

A schematic of the circuit and a detailed description of its parameters are shown in Fig.~\ref{fig:snail-scheme}.

The device parameter space explored is
\[
\mathcal{P}_{\mathrm{SNAIL}} = \left\{ A_J, \rho_{Ic}, \alpha, t, L_\ell, C_\ell, P \right\}
\]
where the parameters are limited by fabrication constraints imposed by lithography and material properties. Their respective ranges are reported in Table~\ref{tab:DevParSpace}.

\begin{table}[h!]
\centering
\begin{tabular}{c||c|c|c|c||c}
\textbf{Parameter} & \textbf{Min} & \textbf{Max} & \textbf{Step} & \textbf{$n_i$} & \textbf{Unit} \\ \hline
$A_J$ & 0.1 & 0.6 & 0.05 & 11 & $\si{\micro\meter}^2$ \\
$\rho_{Ic}$ & 0.5 & 1.5 & 0.1 & 11 & $\si{\micro A/\micro m^2}$ \\
$\alpha$ & 0.23 & 0.25 & 0.02 & 2 & adim. \\
$t$ & 1 & 20 & 1 & 20 & $\mathrm{nm}$ \\
$L_\ell$ & 1.5 & 2 & 0.5 & 2 & adim. \\
$C_\ell$ & 1 & 1.5 & 0.5 & 2 & adim. \\
$P$ & 2 & 3 & 1 & 2 & adim. \\
\end{tabular}
\caption{Values of the device parameters composing $\mathcal{P}_{\mathrm{SNAIL}}$. The column $n_i$ indicates the number of discrete values considered for each parameter.}
\label{tab:DevParSpace}
\end{table}

The simulation is driven by the physical drive quantities:
\[
\mathcal{Q}_{\textrm{SNAIL}} = \left\{ f_{\mathrm{range}}, f_{\textrm{DC}}, A_{\textrm{DC}}^{\mathrm{L}}, A_{\textrm{DC}}^{\mathrm{NL}}, f_{\textrm{p}}, A_{\textrm{p}}^{\mathrm{L}}, A_{\textrm{p}}^{\mathrm{NL}} \right\}
\]
where each source $i$ is characterized by a frequency $f_{\mathrm{i}}$ and amplitudes for linear $A_{\mathrm{i}}^{\mathrm{L}}$ and nonlinear $A_{\mathrm{i}}^{\mathrm{NL}}$ regimes. The simulated frequency span is $f_{\mathrm{range}} = [0, 25] \, \mathrm{GHz}$ with steps of $10 \, \mathrm{MHz}$. The first source (\( f_{\textrm{DC}} = 0 \)) provides a DC flux bias, whose amplitudes depend parametrically on the SNAIL asymmetry \( \alpha \) following a calibration curve \( g_{\mathrm{bias}} \), such that \( A_{\mathrm{DC}}^{\mathrm{L}} = A_{\mathrm{DC}}^{\mathrm{NL}} = g_{\mathrm{bias}}(\alpha) \). This relation enforces 3WM operation, strongly suppressing the four-wave mixing component \cite{nilsson2023}. The second source is a pump tone at \(f_{\textrm{p}} = 11.5 \,\mathrm{GHz} \), with low amplitude in linear simulations and higher amplitude in the nonlinear regime, in our case $A_{\mathrm{p}}^{\mathrm{NL}} = [0.1, 0.5] \, \si{\micro A}$ with range 0.05 $\si{\micro A}$. The expected amplification band is centered in $f_{\mathrm{p/2}} = 5.75 \, \mathrm{GHz}$, considering a bandwidth $f_{\mathrm{BW}} = [4.75, 6.75] \, \mathrm{GHz}$.

The metric function that evaluates the circuit's response in the first two stages is designed to quantify both impedance and phase matching, while also suppressing unwanted harmonic generation:
\begin{equation*}
\mathcal{M}(S(\mathbf{p})) = \frac{\text{a}}{\left|\frac{1}{\Delta f_{\mathrm{BW}}} \int_{f_{\mathrm{BW}}} \mathbf{S}_{11}(f) \, df\right|} + \text{b} \cdot \Delta k + \text{c} \cdot |\mathbf{S}_{11}(f_{\mathrm{2p}})|,
\end{equation*}
where $\text{a}=\text{c}=10$ and $\text{b}=1$ are weighted coefficients empirically optimized. The first term quantifies impedance matching through the mean value of the reflection coefficient $\mathbf{S}_{11}$ at $f_{\mathrm{BW}}$. The second term
\[
\Delta k = |k (f_\mathrm{p}) - 2k(f_\mathrm{p/2})|
\]
evaluates the phase mismatch, where $k(f) = -\operatorname{arg}[\mathbf{S}_{21}(f)]/N$ is the wavenumber at frequency $f$. The third term penalizes transmission at the second harmonic frequency $f_{\mathrm{2p}}$, discouraging second-harmonic generation and promoting stable three-wave mixing operation \cite{nilsson2023}.

\begin{figure}[!t]
\begin{center}
\includegraphics[width=0.4\textwidth]{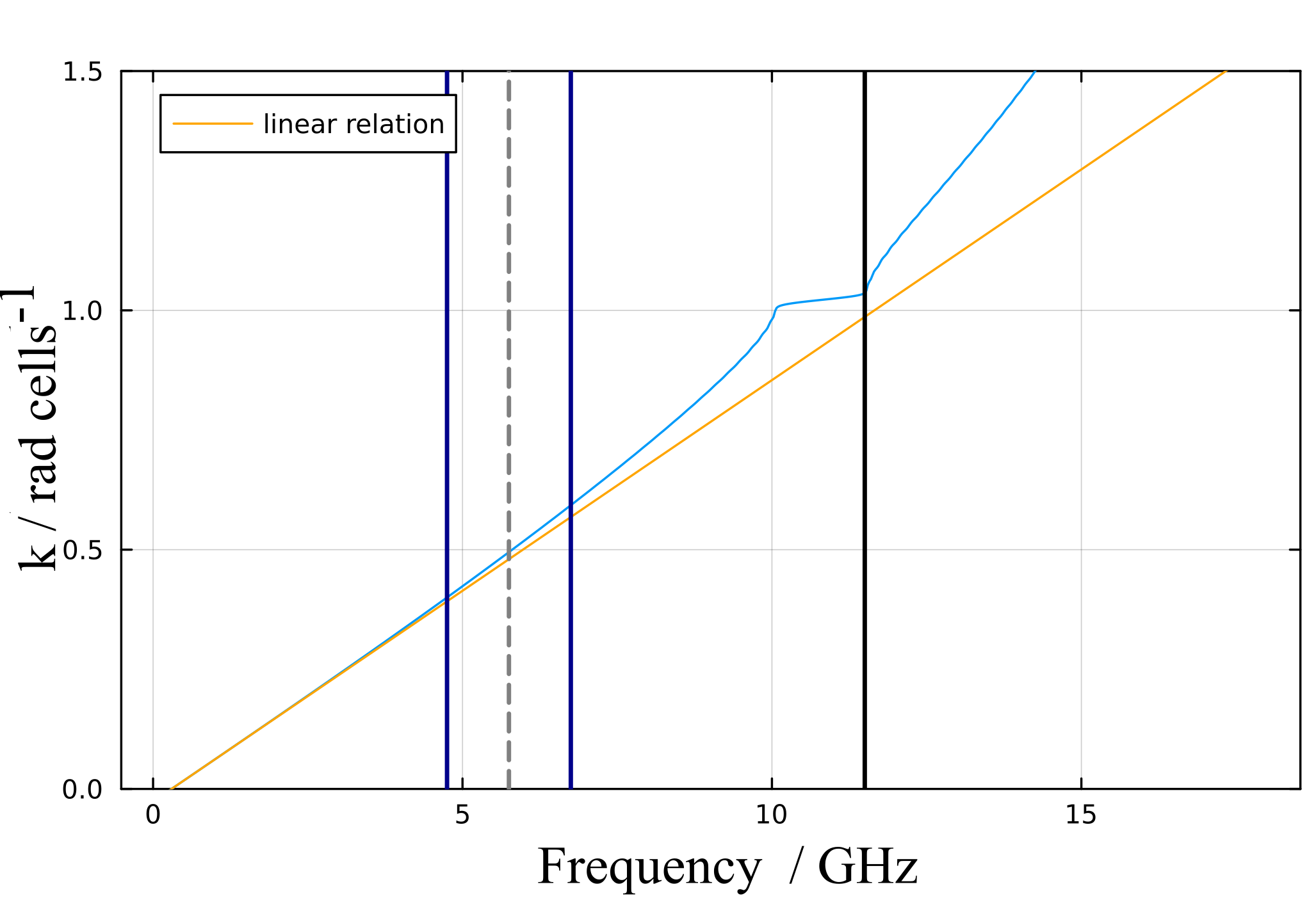}
\end{center}
\caption{Dispersion relation of the optimized SNAIL-based JTWPA, obtained by minimizing the device-specific metric to an optimal circuit configuration $\mathbf{p}^*$. The selected optimal design parameters are $\mathbf{p}^* = \{ A_J = 0.49\,\si{\micro m^2}, \rho_{Ic} = 0.9\,\si{\micro A/\micro m^2}, \alpha = 0.23, t = 9\,\mathrm{nm}, L_\ell = 1.5, C_\ell = 1, P = 3 \}$. The orange line shows the linear relation $k_\mathrm{p} = 2k_\mathrm{p/2}$. Vertical black and dashed lines indicate $f_\mathrm{p}$ and $f_\mathrm{p}/2$, while blue lines define the signal bandwidth $f_\mathrm{BW}$.}
\label{fig:disp}
\end{figure}

\begin{figure}[!t]
\begin{center}
\includegraphics[width=0.4\textwidth]{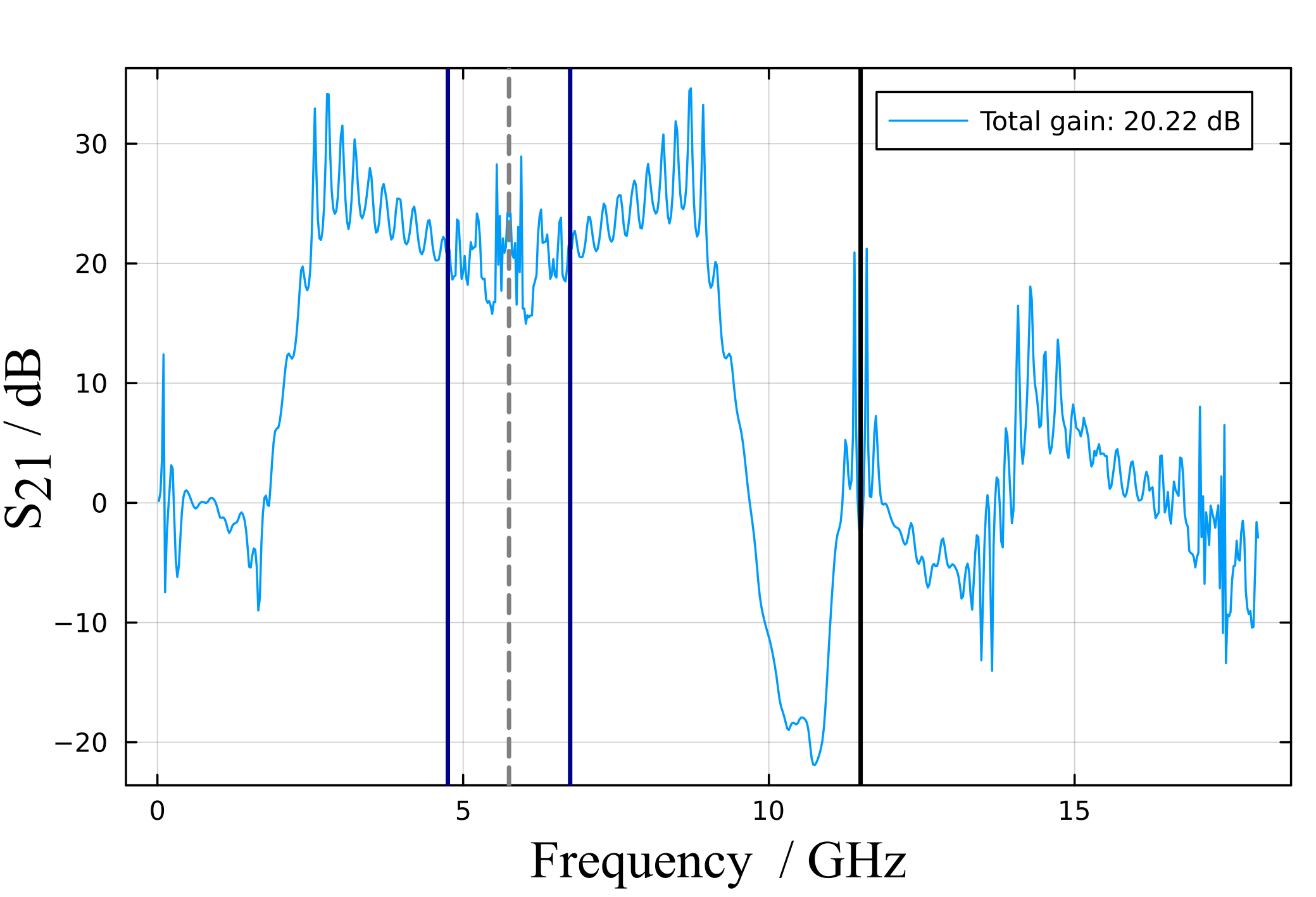}
\end{center}
\caption{Gain profile of the SNAIL-based JTWPA at the optimal working point $\mathbf{q}^*$, obtained by maximizing the performance function $\mathcal{F}(\mathbf{q}, \mathbf{p}^*)$. The selected working point is at $A_{\mathrm{DC}}^{\mathrm{NL}} = 212\,\si{\micro A}$ and $A_{\mathrm{p}}^{\mathrm{NL}} = 0.1\,\si{\micro A}$. At this operating condition, the amplifier achieves a gain of approximately 20\,dB. The vertical lines correspond to those in Fig.~\ref{fig:disp}.}
\label{fig:gain}
\end{figure}

After identifying the optimal design parameters \( \mathbf{p}^* \) via Bayesian optimization, a nonlinear simulation is carried out to determine the optimal working point \( \mathbf{q}^* \). The nonlinear stages assess the circuit performance through
\begin{equation*}
\mathcal{F}\!\left(X(\mathbf{q}; \mathbf{p}^*)\right) = \frac{1}{\Delta f_{\mathrm{BW}}} \int_{f_{\mathrm{BW}}} \mathbf{S}_{21}(f) \, df
\end{equation*}
which quantifies the average transmission gain within the target bandwidth. The optimized configuration achieves a gain of approximately 20 dB at the optimal working point \( \mathbf{q}^* \), as shown in Fig.~\ref{fig:gain}.

The linear simulations require approximately 22 hours, while the Bayesian optimization phase took about 2 hours, exploring a total of $\prod_i n_i$ configurations. Each simulation takes \(\sim2~\mathrm{s}\), dominated by the number of cells and the fine frequency resolution. Simpler devices, such as flux-driven JPAs~\cite{Yamamoto2008}, reduce the runtime to about 0.03 s per simulation. The nonlinear stage tests around 9 different configurations, varying $A_{\textrm{p}}^{\mathrm{NL}}$ and requiring tens of minutes each run. This use case demonstrates how the JCO framework efficiently identifies phase-matched operating regions and optimal drive parameters for complex superconducting circuits, providing a systematic and reproducible design workflow. Here, the reported average gain is obtained without introducing a nonlinear correction term on the linear metric, which accounts for the phase-matching relation; consequently, the gain function contains several ripples. Future work will incorporate the Kerr effect correction, leveraging existing capabilities of the framework to further improve phase matching and amplifier performance.

\section{Conclusion}
\label{ch:4}

In this work, we presented a modular framework for the simulation and optimization of superconducting quantum circuits, combining parametric circuit generation, multi-regime simulations, and Bayesian optimization. Its effectiveness was demonstrated on a SNAIL-based JTWPA in the three-wave mixing regime, where the framework efficiently identified design and operating parameters satisfying impedance matching, phase matching, and broadband gain requirements.

The modular structure allows straightforward extension to other superconducting devices and optimization strategies, enabling systematic exploration of complex circuit topologies. Future developments include implementing nonlinear corrections such as Kerr effects, testing additional superconducting devices, applying dimensionality-reduction techniques like PCA, and exploring alternative optimization strategies. Overall, the framework provides a practical and extensible tool for designing quantum-limited amplifiers and other superconducting quantum technologies.

\section*{Acknowledgments}
This work was partially supported by the European Union’s Horizon Europe Programme (2021--2027) under Grant Agreement No.~101135868 (MiSS), and by the 23FUN08 MetSuperQ project, which received funding from the European Partnership on Metrology, co-financed by the European Union’s Horizon Europe Research and Innovation Programme and the participating states.
\ifCLASSOPTIONcaptionsoff
  \newpage
\fi

\bibliographystyle{IEEEtran}
\bibliography{bibl}

\end{document}